# A Metric for the Activeness of a Class

Sachin Lakra, Deepak Kumar Sharma, Jugnesh Kumar,
Ramesh Chandra Verma, T V Prasad

**Abstract**
In this paper, the authors propose a software metric called Class Activeness Metric (CAM) which helps to determine the level of accessibility of the members of a class when it is instantiated as objects. Object interactions need to be straight forward as far as possible as complexity in these interactions can lead to time delays in accessing members not just confusing inheritance hierarchies. For object interactions to be non-complex, the classes must be designed well so that they are easily accessible. This necessitates the development of a metric for gauging the quality of design of a class. This metric is the Class Activeness Metric.

## 1. Introduction

Object Oriented Technology (OOT) is gained a stronghold in importance and usage in the IT industry as a methodology for development of software. The basis of this technology lies in Component-based Assembly which is prevalent in the world of machines and computer hardware. Pre-designed components are picked off-the-shelf and placed at pre-determined positions to create larger assembled parts, which are further assembled into a complete machine. The same is true of Computer Hardware, where the components are Integrated Circuits and Printed Circuit Boards. In OOT, the components are design time classes and runtime objects. These are "assembled" together in a software module, which are further combined to create a complete software system.

All parts of a machine must take part in the functioning of the complete machine. Each part must function perfectly otherwise the machine will fail at some point or the other. For the parts to function effectively and efficiently, they must be designed well. Each of them must be accessible with easily crossable interfaces and simple internal design. For the design to be considered a good design, its quality of design must be measurable. The CAM is a measure of the accessibility of the members of a class indicating the quality of design of that class.

The paper is divided into sections. Section 1 introduces the paper. Section 2 explains the concept of Activeness. Section 3 defines some basic concepts of Object Oriented Technology, which affect the Class Activeness Metric. Section 4 gives the notation used in the paper. Section 5 considers the effects of visibility of members of a class on the Activeness of the Class. Section 6 explores the effects of Inheritance on the Activeness of a Class. Section 7 reflects on the effect of friend functions on the Activeness of a Class. Section 8 defines the Class Activeness Metric and proposes a mathematical formulation of the CAM. Section 9 concludes the paper.

## 2. Activeness
### 2.1 A Recent Concept.
Activeness is a recent new concept and is defined as the degree of readiness of a system to respond to the stimuli from the environment in which it exists [1]. Activeness may be related to any living entity or any system which exists in any environment in the universe. The Activeness of vacuum without any stimuli and no system existing in it is zero [1].

**2.2 Why does Activeness exist in a system?**
The question arises as to why a given system possesses Activeness. The answer is that every system has some "Organizedness" in it, that is, there is some degree of order in the system. The definition of a system itself says that "a System is a set of components working together to achieve a goal." Components cannot work together if they are not organized. This, in turn, implies that if a system is given an external stimulus there will be some change in the degree of order of the system, that is, there will be a response of the system to the stimulus. Every system responds to such stimuli. The "Organizedness" of the system makes it ready to respond to them to some degree. This degree of readiness of a system to respond to a stimulus is the concept called Activeness [3].

**2.3 What is the need to study the Activeness of a system?**

Another important question is that why Activeness should be studied at all. The answer lies in the fact that the observer wants to know how far she can depend on a system and its response. The system can respond well if it has the readiness to do so and if it is stable. It is stable if it is organized, i.e., if it has order. Thus if the Activeness of a system is known, its stability and how well it can respond to the stimulus it will be given, can be judged[3].

**2.4 Origin of the Concept of Activeness**

The first author of this paper made the observation that if a human being tries to ward off a honeybee by waving a hand at it, the honeybee may get angered and may bite the human being. He wondered why the honeybee, being so small, is able to hurt a much bigger human being. The obvious answer was that the honeybee is *capable* of biting the human being and is therefore able to bite him. Further, the honeybee was capable because it was a system comprised of various biological subsystems, one of which was the sting, that is, the honeybee was very well organized and was ready to defend itself in case of an attack. This led to the idea of studying various systems as to how they can react to environmental stimuli.

**3. Classes and Objects**
  **3.1. What is a class?**
A class is an encapsulation of high-level abstractions of real-life entities, and exhibits the property of data hiding. The major action that can be performed by or on a class(es) is inheritance. A class exists only at design time.
  **3.2. What is an object?**
An object is an instance of a class. In object oriented programming, an object is a class variable. An object exists only at run-time.
  **3.3. Structure of a class**

A class encapsulates the two components of data and functionality. Data is in the form of member variables (or properties or attributes) and functionality is in the form of member functions (or methods or messages).

### 3.4. Types of visibility in a class

As defined in most Object Oriented languages, a class allows three types of visibility, namely, private, public and protected, for its data and functionality.

### 3.5. Inheritance

Inheritance is the concept of derivation of the data or functionality of a parent class by a new class, which is similar to the parent class but has certain specialized features of its own.

## 4. Notation used

The notation used in Table 1 below and the rest of this paper is summarized as follows:
C = Class under consideration,
$C_A$ = Activeness of Class C,
$C_{AN}$ = Activeness of Class C under no inheritance,
$C_{AU}$ = Activeness of Class C under public inheritance,
$C_{AI}$ = Activeness of Class C under private inheritance,
$C_{AR}$ = Activeness of Class C under protected inheritance,
$n_i$ = number of levels of inheritance,
$n_f$ = number of friend functions in a class,

$v_u$ = number of public member variables,
$f_u$ = number of public member functions,
$v_i$ = number of private member variables,
$f_i$ = number of private member functions,
$v_r$ = number of protected member variables,
$f_r$ = number of protected member functions,

$v_{ui}$ = sum of number of member variables publicly inherited from class C by all the derived classes or by C from all the base classes,
$f_{ui}$ = sum of number of member functions publicly inherited from class C by all the derived classes or by C from all the base classes,
$v_{ii}$ = sum of number of member variables privately inherited from class C by all the derived classes or by C from all the base classes,
$f_{ii}$ = sum of number of member functions privately inherited from class C by all the derived classes or by C from all the base classes,
$v_{ri}$ = sum of number of member variables inherited from class C by all the derived classes or by C from all the base classes under protected inheritance,
$f_{ri}$ = sum of number of member variables inherited from class C by all the derived classes or by C from all the base classes under protected inheritance,

$n_{lu}$ = number of levels of public inheritance,
$n_{li}$ = number of levels of private inheritance,
$n_{lr}$ = number of levels of protected inheritance.

## 5. Effects of visibility on the Activeness of a Class.

The Activeness of a class depends on the visibility levels of each of the various members of the class. This dependence is described below.

**5.1** The Activeness $C_A$, of a class C depends directly on the sum $s_u$ of the number of public member variables $v_u$ and the number of public member functions $f_u$ in the class, i.e.,

$$C_A \propto s_u = v_u + f_u \qquad (1)$$

**5.2** The Activeness $C_A$, of a class C depends inversely on the sum $s_i$ of the number of private member variables $v_i$ and the number of private member functions $f_i$ in the class, i.e.,

$$C_A \propto \frac{1}{s_i} = \frac{1}{v_i + f_i} \qquad (2)$$

**5.3** The Activeness $C_A$, of a class C depends directly on the sum $s_r$ of the number of protected member variables $v_r$ and the number of protected member functions $f_r$ in the class, i.e.,

$$C_A \propto s_r = v_r + f_r \qquad (3)$$

### 5.4 Activeness of a Class with no inheritance.

The case when there is no inheritance, where the various public and protected member functions can access the rest of the members of the class C under consideration, is a special case. Here, the greater the number of public and protected member variables and member functions defined within the class, greater the accessibility of C. But, the greater the number of private members variables and member functions in C, lesser the accessibility of C. This is represented by the following equation:

$$C_A \propto C_{AN} = \frac{(s_u + s_r)}{s_i}, n_i = 0 \qquad (4)$$

## 6 Effects of Inheritance on the Activeness of a Class.

The Activeness $C_A$, of a class C, depends on two parameters of inheritance, namely, the sum of the number of member variables and the number of member functions inherited from C or by C, and the number of levels of inheritance $n_l$, according to the visibility levels.

This dependency of $C_A$ on the type of visibility in inheritance for each of the two parameters mentioned above is shown in the following table and explained below:

| Parameter | Inheritance | | |
| --- | --- | --- | --- |
| | Public | Private | Protected |
| Sum of the number of member variables inherited and the number of member functions inherited | $C_A \propto i_u$, where $i_u = v_{ui} + f_{ui}$ | $C_A \propto \frac{(i_u + i_r)}{i_i}, n_{li} = 1$, where $i_i = v_{ii} + f_{ii}$, $i_u = v_{ui} + f_{ui}$ and $i_r = v_{ri} + f_{ri}$ | $C_A \propto (i_u + i_r), n_{lr} = 1$, where $i_r = v_{ri} + f_{ri}$ |

| Number of levels of inheritance, $n_l$ | $C_A \propto n_{lu}$ | Defined only for $n_{li}=1$, where $n_{li}$ does not affect $C_A$ | $C_A \propto n_{lr}, n_{lr}=1$ |

Table 1: Dependency of the Activeness $C_A$ of a class on inheritance parameters and the type of visibility.

### 6.1 Public Inheritance

In Public inheritance, the greater the number of member variables and member functions inherited from class C, greater the accessibility of C. Similarly, greater the number of levels of public inheritance, greater is the accessibility of the members of the Class C under consideration and hence, greater the Activeness of C. From Table 1, the following relation is obtained for Public inheritance:

$$C_A \propto C_{AU} = n_{lu} \times i_u \quad (5)$$

where
$i_u = v_{ui} + f_{ui}$

### 6.2 Private Inheritance

In Private inheritance, the case of the number of levels being 1, i.e., where $n_{li}=1$, is a special case. This case is to be considered since public and protected members inherited privately become private in the derived class. Thus, the greater the sum of the number of public and protected member variables and member functions privately inherited from class C, greater the accessibility of C. Also, the greater the sum of the number of private member variables and member functions privately inherited from class C, lesser the accessibility of C, as private members are not inherited in this type of inheritance. From Table 1, the following relation is obtained for Private inheritance up to one level:

$$C_A \propto C_{AI} = \left[\frac{(i_u + i_r)}{i_i}\right], n_{li}=1, i_i > 0 \quad (6)$$

where $i_u = v_{ui} + f_{ui}$, $i_r = v_{ri} + f_{ri}$, $i_i = v_{ii} + f_{ii}$.

For the rest of the cases, where $n_{li}>1$, there is no inheritance possible in the case of Private inheritance.

### 6.3 Protected Inheritance

In Protected inheritance also, the case of $n_{lr}=1$ is special. In this case, the public and protected members inherited from a base class by a class C under consideration, become private in class C and cannot be inherited further. Here, the greater the sum of the numbers of protected or public members in a base class, greater will be the accessibility of C and hence, greater the Activeness of C. For number of levels of inheritance greater than 1, the number of publicly inheritable members under protected inheritance will become zero, as both protected and public members are inherited as protected members by a second level derived class. From Table 1, the following relation is obtained for Protected inheritance:

$$C_A \propto C_{AR} = \left[\frac{n_{lr} \times (i_u + i_r)}{i_i}\right], n_{lr} \geq 1 \qquad (7)$$

where, $i_u = v_{ui} + f_{ui}$ and $i_r = v_{ri} + f_{ri}$

## 7 Effect of friend functions on the Activeness of a Class.

The Activeness $C_A$, of a Class, depends on the number of friend functions, $n_f$, accessing the members of the class. Visibility does not have any effect on the friend functions accessing a class. Thus,

$$C_A \propto n_f \qquad (8)$$

## 8 Class Activeness Metric(CAM):

CAM is defined as the level of readiness of a class to be accessed by its own member functions or by member functions of another class. The Activeness of a class, $C_A$, can be measured by the Class Activeness Metric by the following:

$$CAM = \frac{(s_u + s_r)}{s_i} + n_f, n_i = 0 \qquad (9a)$$

$$CAM = \left[\left\{(n_{lu} \times i_u) + \frac{(n_{lr} + 1) \times (i_u + i_r)}{i_i}\right\} + n_f\right], \qquad (9b)$$

where $n_i > 0, i_i > 0$.

or

$$CAM = C_{AN} + n_f, n_i = 0 \qquad (10a)$$

$$CAM = C_{AU} + C_{AR} + C_{AI} + n_f, n_i > 0 \qquad (10b)$$

### 8.1.1 Advantages and limitations

The advantages of the CAM include
- The CAM provides a means of gauging the accessibility of a class.
- The CAM acts as an indicator of the quality of design of a class.
- A cumulative CAM of all the classes and inheritance hierarchies of a single module of a software system could be used as a measure of the quality of design of the module.
- A cumulative CAM of all the classes in all the modules of the software system taken together can be developed as a metric for the quality of design of the complete software system.

A limitation of the CAM is that the CAM should be considered only as an indication of the quality of design of a Class. However, this implies that there will be a tradeoff between accessibility and security as implied by data hiding in the process of an attempt to improve the design of a class by improving the value of the CAM.

### 8.1.2 Need of the CAM

The design of a software system is a major issue in the correct and successful functioning of that system upon implementation. Presently, hardly any measures exist for measuring the quality of design of a software system. The CAM completes this unfulfilled requirement.

## 9 Conclusion

The CAM is thus an indicator of the quality of design of a class in any object oriented system. The CAM also indicates, in particular, that to increase the accessibility of private member functions in a class, more friend functions should be declared in the class, although this conflicts with the concept of data hiding and security. The shortcoming of the CAM is that it is a theoretical concept which has not been practically implemented at present. The authors intend to test it by applying it on classes of an object oriented system.

## 10 References

"A Metric For The Activeness Of An Object-Oriented Component Library", Proceedings of Software Engineering Research & Practice (SERP'07), WORLDCOMP '07 Conference, Las Vegas, Nevada, USA; $25^{th}$ – $28^{th}$ June 2007; pp. 704-709; by **Sachin Lakra**, Nand Kumar, Sugandha Hooda, Nitin Bhardwaj.

"Metrics For The Pre-Development Phase Of Software Requirements Engineering"; Proceedings (Abstract) of National Conference on Emerging Trends in Software Engineering and Information Technology, (Editor and Convener: Prof Dr.Hemant K. Soni), held at Gwalior Engineering College, Gwalior, M.P., India; $29^{th}$ – $30^{th}$ March, 2007; pp. 21; by **Sachin Lakra**, Bharti Jha, Nitin Bhardwaj, Ritu Saluja and Nand Kumar.

Object Oriented Programming with C++, by E Balagurusamy, Third Edition, 2006, Tata McGraw Hill Publishing Company Limited, New Delhi.

http://www.cse.iitb.ac.in/~rkj/617-08/lecture8.pdf